\documentclass[aps,prx,reprint,amsmath,amssymb,superscriptaddress]{revtex4-2}

\usepackage[utf8]{inputenc}
\usepackage{graphicx}
\usepackage{booktabs}
\usepackage{hyperref}
\usepackage{xcolor}
\usepackage{algorithmicx}
\usepackage{algpseudocode}
\usepackage{tikz}
\usepackage{amsthm}

\usetikzlibrary{calc,patterns,decorations.pathmorphing}

\begin{document}

\title{Stochastic Loop Corrections to Belief Propagation for Tensor Network Contraction}

\author{Gi Beom Sim}
\thanks{Contributed equally to this work}
\affiliation{Department of Energy Science, Sungkyunkwan University, Seobu-ro 2066, Suwon, 16419, Korea}

\author{Tae Hyeon Park}
\thanks{Contributed equally to this work}
\affiliation{Department of Energy Science, Sungkyunkwan University, Seobu-ro 2066, Suwon, 16419, Korea}

\author{Kwang S. Kim}
\affiliation{Department of Chemistry, Ulsan National Institute of Science and Technology, 50 UNIST-gil, Ulsan 44919, Republic of Korea}

\author{Yanmei Zang}
\affiliation{Department of Energy Science, Sungkyunkwan University, Seobu-ro 2066, Suwon, 16419, Korea}

\author{Xiaorong Zou}
\affiliation{Department of Energy Science, Sungkyunkwan University, Seobu-ro 2066, Suwon, 16419, Korea}

\author{Hye Jung Kim}
\email{hjkim75@skku.edu}
\affiliation{Department of Energy Science, Sungkyunkwan University, Seobu-ro 2066, Suwon, 16419, Korea}

\author{D. ChangMo Yang}
\email{dcyang@skku.edu}
\affiliation{Department of Energy Science, Sungkyunkwan University, Seobu-ro 2066, Suwon, 16419, Korea}

\author{Soohaeng Yoo Willow}
\email{sy7willow@gmail.com}
\affiliation{Department of Energy Science, Sungkyunkwan University, Seobu-ro 2066, Suwon, 16419, Korea}

\author{Chang Woo Myung}
\email{cwmyung@skku.edu}
\affiliation{Department of Energy Science, Sungkyunkwan University, Seobu-ro 2066, Suwon, 16419, Korea}
\affiliation{Department of Quantum Information Engineering, Sungkyunkwan University, Seobu-ro 2066, Suwon, 16419, Korea}
\affiliation{Department of Energy, Sungkyunkwan University, Seobu-ro 2066, Suwon, 16419, Korea}

\date{\today}

% Abstract
% Abstract
\begin{abstract}
Tensor network contraction is a fundamental computational challenge underlying quantum many-body physics, statistical mechanics, and machine learning. Belief propagation (BP) provides an efficient approximate solution, but introduces systematic errors on graphs with loops. Here, we introduce a hybrid method that achieves accurate results by stochastically sampling loop corrections to BP and showcase our method by applying it to the two-dimensional ferromagnetic Ising model. For any pairwise Markov random field with symmetric edge potentials, our approach exploits an exact factorization of the partition function into the BP contribution and a loop correction factor summing over all valid loop configurations, weighted by edge weights derived directly from the potentials. We sample this sum using Markov chain Monte Carlo with moves that preserve the loop constraint, combined with umbrella sampling to ensure efficient exploration across all correlation strengths. Our stochastic approach provides unbiased estimates with controllable statistical error in any parameter regime.
\end{abstract}

\maketitle

% Main sections
% Introduction
\section{Introduction}
\label{sec:introduction}

Tensor network contraction is a fundamental computational task with broad applications across physics, chemistry, and computer science~\cite{orus2014practical,cirac2021matrix}. In quantum many-body physics, tensor networks provide compact representations of quantum states, including matrix product states (MPS) for one-dimensional systems~\cite{verstraete2008matrix,perezgarcia2007matrix,chan2011Density} and projected entangled pair states (PEPS) for higher dimensions~\cite{cirac2021matrix,Wahl2025}, where physical observables are computed by contracting the network of tensors. In statistical mechanics, the partition function of classical spin models can be expressed as a tensor network, with contraction yielding thermodynamic quantities~\cite{nishino1996corner}. In machine learning, tensor network architectures have been applied to supervised learning~\cite{stoudenmire2016supervised}, generative modeling~\cite{han2018unsupervised}, and dimensionality reduction~\cite{cichocki2016tensor}. The common computational bottleneck across all these applications is tensor network contraction, which involves computing the resulting scalar or lower-rank tensor from a network of tensors connected by contracted indices.

Unfortunately, exact tensor network contraction is generically \#P-hard~\cite{schuch2007computational}, placing it beyond the reach of efficient classical algorithms for general networks. The computational cost scales exponentially with the treewidth of the underlying graph, making exact contraction infeasible for two-dimensional and higher-dimensional networks of even modest size. Consequently, approximate contraction methods are essential for practical applications ranging from variational optimization~\cite{verstraete2008matrix,perezgarcia2007matrix} to combinatorial optimization~\cite{kourtis2019fast} and quantum circuit simulation~\cite{markov2008simulating,gray2021hyper}.

Belief propagation (BP) has emerged as a powerful approximate contraction algorithm~\cite{pearl1988probabilistic,mezard2009information,alkabetz2021tensor}. Originally developed for probabilistic inference on graphical models, BP operates by iteratively passing ``messages'' along network edges until convergence. Each message encodes the marginal probability distribution of a variable given information from its neighbors, and the algorithm exploits the local structure of the network to achieve global inference. On tree-structured networks, which are graphs without cycles, BP converges it to the exact marginal distributions and partition function in time linear in the number of edges~\cite{pearl1988probabilistic}. This remarkable efficiency has made BP a cornerstone of probabilistic inference~\cite{kschischang2001factor,wainwright2008graphical}, error-correcting codes~\cite{richardson2001design}, and constraint satisfaction problems~\cite{mezard2002analytic}.

On graphs with loops, BP no longer provides exact results but instead computes the Bethe approximation~\cite{yedidia2003understanding,yedidia2005constructing}. The Bethe free energy assumes that correlations between variables are purely local, captured by pairwise interactions along edges without accounting for higher-order correlations that propagate around cycles. When BP converges on a loopy graph, the fixed point corresponds to a stationary point of this Bethe free energy~\cite{yedidia2003understanding}. The computational efficiency of BP, which still scales linearly with the number of edges, has made it an attractive approximation for tensor network contraction~\cite{jiang2008accurate,park2025simulating}, particularly for large-scale problems where exact methods are infeasible. 

Despite its computational efficiency, BP introduces systematic errors on graphs with loops. This limitation arises because BP assumes that neighboring nodes are statistically independent, an assumption that holds only for loop-free (tree-like) network structures. Since virtually all realistic models contain loops, standard message passing can yield poor results in practice~\cite{heskes2004uniqueness,mooij2007loop,kirkley2021belief}. In a loop, BP treats each edge independently, so correlation information traveling around the loop returns to reinforce itself, biasing the partition function and marginals. These errors remain small in weakly correlated regimes but become substantial near phase transitions or in frustrated systems~\cite{ricci2012bethe,dominguez2011characterizing}.

Recognizing these limitations, recent work has developed systematic corrections to BP that attempt to account for loop contributions~\cite{gray2025tensor,evenbly2026loop}. Loop expansions, pioneered by Chertkov and Chernyak~\cite{chertkov2006loop,chertkov2008loop}, express the exact partition function as the BP result multiplied by a series over generalized loops, which are subgraphs where every vertex has even degree. Each term in this ``loop series'' can be computed from BP messages, providing a systematic correction hierarchy. TAP (Thouless-Anderson-Palmer) corrections~\cite{thouless1977solution,plefka1982convergence,opper2001adaptive} provide closed-form second-order corrections that capture the reaction field effect. Cluster expansions~\cite{midha2025beyond} sum contributions from connected clusters of variables to handle overlapping contributions. In the context of PEPS, the Simple Update algorithm~\cite{jiang2008accurate} uses an environment approximation equivalent to BP's Bethe approximation. Lubasch, Cirac, and Bañuls~\cite{lubasch2014unifying,lubasch2014algorithms} introduced a cluster update strategy that systematically interpolates between this Simple Update and the full environment contraction by treating clusters of size $\delta$ around each tensor exactly, demonstrating that the contraction error decreases exponentially with $\delta$ at a rate governed by the correlation length of the state.

However, these analytical correction methods share fundamental limitations. Loop series can diverge precisely in the strongly correlated regime where corrections are most needed~\cite{ihler2005loopy,mooij2007sufficient}. Near phase transitions, where correlations become long-range and critical fluctuations dominate, the series often fails to converge. TAP corrections, while stable, are limited to second order and become inaccurate for strong correlations. None of these methods provides a systematic, convergent path from the Bethe approximation to the exact result across all parameter regimes.

Recent advances have explored alternative directions to improve inference on loopy graphs. Hyperoptimized tensor network contraction~\cite{gray2024hyper} employs bond compression with automatic hyperparameter tuning over contraction strategies, achieving orders of magnitude speedup for approximate contraction. Graph neural networks have been trained to learn message-passing algorithms that outperform BP on loopy graphs~\cite{kuck2020gnn}, with demonstrated out-of-distribution generalization to larger systems. A unified framework for BP on graphs with loops~\cite{hack2025belief} shows that block BP and tensor network message passing are special cases of a general construction, enabling systematic improvements. Stochastic tensor contraction~\cite{sun2026stochastic} has recently been introduced for quantum chemistry, where importance sampling over tensor indices reduces contraction cost for coupled cluster calculations.

In this work, we introduce a fundamentally different approach based on stochastic sampling of loop configurations via Markov chain Monte Carlo (MCMC). Rather than summing analytical corrections that may diverge, we directly sample the loop configurations appearing in the exact high-temperature expansion of the partition function. For any pairwise Markov random field with symmetric edge potentials $\psi(s_i, s_j)$, this expansion expresses the partition function as the BP result multiplied by a sum over all valid loop configurations, weighted by products of edge weights $u_e$. Each loop configuration is a subgraph where every vertex has even degree. By sampling these configurations stochastically, we avoid convergence issues entirely. The Monte Carlo estimator is unbiased regardless of correlation strength, with accuracy controlled only by sampling statistics.

This approach is inspired by diagrammatic Monte Carlo methods in quantum many-body theory~\cite{prokofev1998polaron,kozik2010diagrammatic,wang2025variational}, where Feynman diagrams contributing to Green's functions are sampled stochastically rather than summed analytically. Just as diagrammatic Monte Carlo can access strong-coupling regimes where perturbation theory diverges, our loop Monte Carlo can access strong-correlation regimes where analytical loop series could fail. The key insight is that while the number of loop configurations grows exponentially with system size, Monte Carlo importance sampling can efficiently explore this space, focusing computational effort on configurations that contribute most to the partition function.

We demonstrate this Belief Propagation Loop Monte Carlo (BPLMC) approach on the two-dimensional ferromagnetic Ising model. The partition function of the Ising model can be expressed as a tensor network contraction, where each vertex hosts a tensor encoding local Boltzmann weights and computing $Z$ requires contracting all tensor indices. On small lattices where exact enumeration is feasible, BPLMC matches exact results to within statistical precision, validating the method. On larger lattices, we compare against the Onsager exact solution~\cite{onsager1944crystal} and demonstrate substantial improvement over BP across all temperatures, with the largest gains in the low-temperature regime where BP errors are most severe.

% Theory
\section{Theory}
\label{sec:theory}

BP is a general message-passing algorithm for probabilistic inference on graphical models~\cite{pearl1988probabilistic,kschischang2001factor,yedidia2003understanding}. The Ising model and other pairwise Markov random fields (MRFs) are a natural subset of this general framework. Our stochastic loop correction method exploits properties specific to pairwise MRFs with symmetric edge potentials, where the high-temperature expansion takes a particularly simple form.

We consider pairwise MRFs defined on a graph $G = (V, E)$ with discrete variables $\{s_i\}_{i \in V}$. The joint probability distribution factorizes as
\begin{equation}
    P(\mathbf{s}) = \frac{1}{Z} \prod_{i \in V} \phi_i(s_i) \prod_{(i,j) \in E} \psi_{ij}(s_i, s_j),
    \label{eq:mrf}
\end{equation}
where $\phi_i(s_i)$ are node potentials, $\psi_{ij}(s_i, s_j)$ are edge potentials, and $Z = \sum_{\mathbf{s}} \prod_i \phi_i \prod_{ij} \psi_{ij}$ is the partition function.

For pairwise MRFs, BP computes approximate marginals by iteratively passing messages along edges. Messages $m_{i \to j}(s_j)$ from vertex $i$ to neighbor $j$ satisfy the update equations:
\begin{equation}
    m_{i \to j}(s_j) \propto \sum_{s_i} \phi_i(s_i) \psi_{ij}(s_i, s_j) \prod_{k \in \partial i \setminus j} m_{k \to i}(s_i),
    \label{eq:bp_message}
\end{equation}
where $\partial i$ denotes the neighbors of $i$. At convergence, the \textit{node beliefs} $b_i(s_i)$ and \textit{edge beliefs} $b_{ij}(s_i, s_j)$ are
\begin{eqnarray}
    b_i(s_i) & \propto & \phi_i (s_i) \prod_{j \in \partial i} m_{j \to i} (s_i), \\
    b_{ij} (s_i, s_j) & \propto & \phi_i(s_i) \phi_j(s_j) \psi_{ij} (s_i, s_j) \nonumber \\
    & & \prod_{k \in \partial i \setminus j} m_{k \to i }(s_i) \prod_{k \in \partial j \setminus i} m_{k \to j} (s_j).
\end{eqnarray}
The Bethe approximation to the partition function~\cite{yedidia2003understanding} is
\begin{equation}
    \log Z_{\mathrm{BP}} = \sum_{(i,j) \in E} \log Z_{ij} - \sum_{i \in V} (d_i - 1) \log Z_i,
    \label{eq:bethe_logZ}
\end{equation}
where $d_i = |\partial i|$ is the degree of vertex $i$.
The quantities $Z_i$ and $Z_{ij}$ represent local partition functions.
They are computed using converged BP messages together with  the corresponding node and edge potentials.
These local terms are defined as
\begin{eqnarray}
    \log Z_{ij} & = & \sum_{s_i, s_j} b_{ij} \log \frac{\psi_{ij} \phi_i \phi_j} {b_{ij}},  \\
    \log Z_i & = & \sum_{s_i} b_i \log \frac{\phi_i}{b_i}.
\end{eqnarray}

The loop expansion requires two conditions on the pairwise MRF: (1) each variable is binary, taking values $s_i \in \{+1, -1\}$, and (2) the edge potentials must be symmetric, such that $\psi_{ij}(+1,+1) = \psi_{ij}(-1,-1) \equiv \psi_{\mathrm{same}}$ and $\psi_{ij}(+1,-1) = \psi_{ij}(-1,+1) \equiv \psi_{\mathrm{diff}}$.
In the absence of an external field, the node potential is uniform, i.e., $\phi_i(s_i) = 1$.

Expanding the product over all edges and using the identity $\sum_{s_i = \pm 1} s_i^{n} = 2\delta_{n \, \mathrm{even}}$, we obtain the exact high-temperature (loop) expansion,
\begin{equation}
    Z = Z_{\mathrm{norm}} \sum_{G \in \mathcal{L}} \prod_{e \in G} u_e,
    \label{eq:high_T_general}
\end{equation}
where $\mathcal{L}$ is the set of valid loop configurations (subgraphs where every vertex has even degree), and the normalization factor is:
\begin{equation}
    Z_{\mathrm{norm}} = \prod_{i \in V} \left(\sum_{s_i} \phi_i(s_i)\right) \prod_{(i,j) \in E} \frac{\psi_{\mathrm{same}} + \psi_{\mathrm{diff}}}{2}.
    \label{eq:Z_norm}
\end{equation}

For uniform node potentials $\phi_i = 1$ and uniform edge potentials (all edges have the same $u_e = u$), this simplifies to:
\begin{equation}
    Z = Z_{\mathrm{BP}} \sum_{G \in \mathcal{L}} u^{|G|},
    \label{eq:high_T_uniform}
\end{equation}
where $Z_{\mathrm{BP}} = 2^N \left(\frac{\psi_{\mathrm{same}} + \psi_{\mathrm{diff}}}{2}\right)^{|E|}$ is the Bethe approximation corresponding to the paramagnetic fixed point, and $|G|$ counts the edges in configuration $G$.

For the ferromagnetic Ising model with Hamiltonian $H = -J \sum_{\langle ij \rangle} s_i s_j$ ($J > 0$), the normalization becomes $Z_{\mathrm{BP}} = 2^N \cosh(\beta J)^{|E|}$, recovering the well-known Bethe approximation for the Ising model. The partition function is thus:
\begin{equation}
    Z = 2^N \cosh(\beta J)^{|E|} \sum_{G \in \mathcal{L}} \tanh(\beta J)^{|G|}.
    \label{eq:ising_Z}
\end{equation}

We define the loop correction factor as
\begin{equation}
    Z_{\mathrm{loop}} = \sum_{G \in \mathcal{L}} \prod_{e \in G} u_e = 1 + \sum_{G \neq \emptyset} \prod_{e \in G} u_e,
    \label{eq:Z_loop}
\end{equation}
so that $Z = Z_{\mathrm{BP}} \cdot Z_{\mathrm{loop}}$ (for the uniform edge weight case). The empty graph ($G = \emptyset$) contributes unity, corresponding to the BP approximation. Non-empty loop configurations provide corrections that become increasingly important as $|u| \to 1$.

Computing $Z_{\mathrm{loop}}$ exactly requires summing over exponentially many loop configurations. However, since each term $\prod_{e \in G} u_e$ is non-negative for ferromagnetic systems ($u > 0$), we can interpret the normalized weights as a probability distribution over loop configurations and estimate $Z_{\mathrm{loop}}$ via Monte Carlo sampling. The key algorithmic challenge is to design MCMC moves that efficiently explore the space of valid loop configurations while preserving the even-degree constraint.

% Methods
\section{Methods}
\label{sec:methods}

%\subsection{Belief propagation implementation}
%\label{subsec:bp_impl}

We implement BP following the formulation of Refs.~\cite{alkabetz2021tensor,yedidia2003understanding}. Messages are initialized uniformly as $m_{i \to j}(s) = 1/n_{\mathrm{states}}$ for all states $s$, ensuring convergence to the paramagnetic fixed point. The Bethe free energy is computed from converged messages as Eq.~(\ref{eq:bethe_logZ}). For MCMC sampling, we construct a cycle basis for the graph using elementary plaquettes. On an $L \times L$ square lattice with periodic boundary conditions, elementary plaquettes are the $L^2$ unit squares. For the cycle basis, we use $L^2 - 1$ independent plaquettes plus two winding cycles around the unit cell, giving $L^2 + 1$ total cycles, matching the cycle space dimension $|E| - |V| + 1$. The cycle basis defines allowed MCMC moves, where any symmetric difference of the current configuration $G$ with a basis cycle preserves the even-degree constraint.

We sample the loop partition function $Z_{\mathrm{loop}}$ using MCMC with moves that preserve the even-degree constraint. The sampling employs key techniques, including plaquette flip moves, multi-cycle moves, winding cycle moves, and umbrella sampling. And statistical errors are estimated via block averaging~\cite{flyvbjerg1989error}.

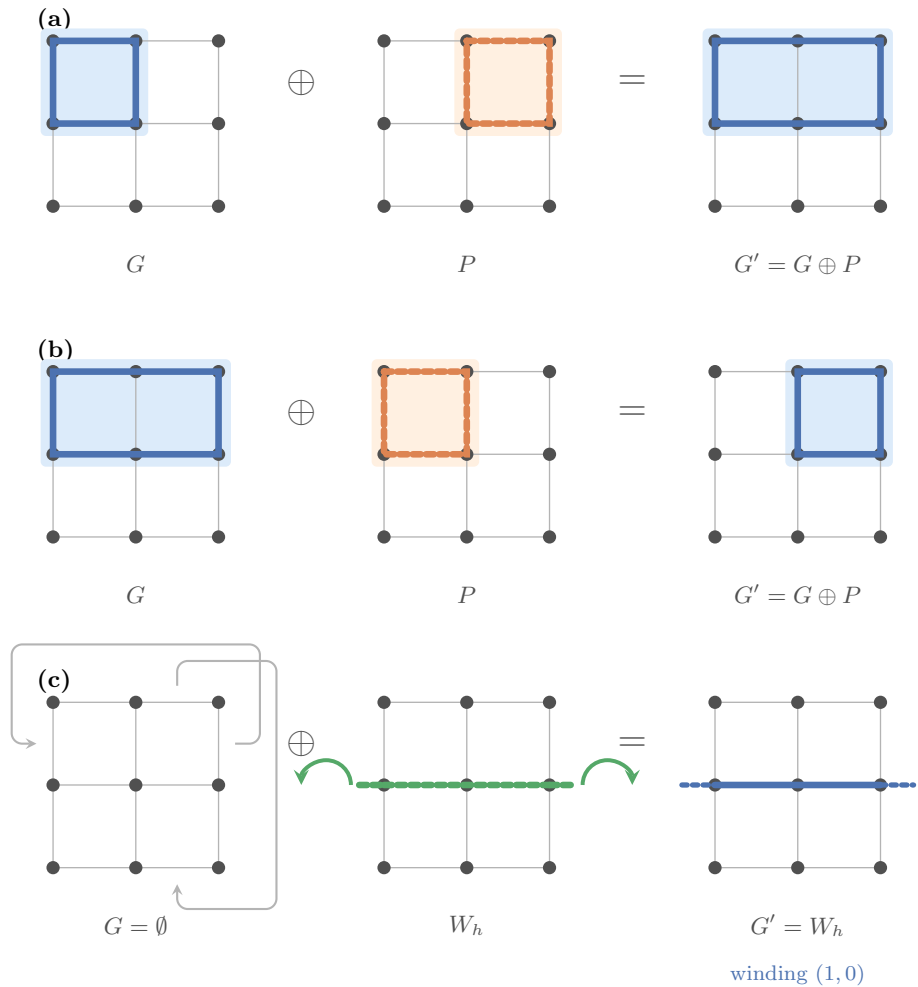
\begin{figure*}[tb]
\centering
\begin{tikzpicture}[scale=1.1, every node/.style={scale=1.0}]
    % Define colors matching seaborn deep palette
    \definecolor{loopblue}{RGB}{76,114,176}    % seaborn blue
    \definecolor{plaqorange}{RGB}{221,132,82}  % seaborn orange
    \definecolor{windgreen}{RGB}{85,168,104}   % seaborn green
    \definecolor{gridgray}{RGB}{180,180,180}
    \definecolor{nodegray}{RGB}{80,80,80}
    \definecolor{bgblue}{RGB}{220,235,250}
    \definecolor{bgorange}{RGB}{255,240,225}

    % Style definitions
    \tikzstyle{loopedge}=[loopblue, line width=2.5pt, line cap=round]
    \tikzstyle{plaqedge}=[plaqorange, line width=2.5pt, line cap=round, densely dashed]
    \tikzstyle{windedge}=[windgreen, line width=2.5pt, line cap=round, densely dashed]
    \tikzstyle{gridedge}=[gridgray, line width=0.5pt]
    \tikzstyle{sitenode}=[circle, fill=nodegray, inner sep=0pt, minimum size=5pt]
    \tikzstyle{opnode}=[font=\Large\bfseries, text=nodegray]
    \tikzstyle{labelnode}=[font=\small, text=nodegray]

    % Helper macro for drawing grid
    \newcommand{\drawgrid}{
        % Draw light grid edges
        \foreach \x in {0,1} {
            \foreach \y in {0,1,2} {
                \draw[gridedge] (\x,\y) -- (\x+1,\y);
            }
        }
        \foreach \x in {0,1,2} {
            \foreach \y in {0,1} {
                \draw[gridedge] (\x,\y) -- (\x,\y+1);
            }
        }
        % Draw nodes on top
        \foreach \x in {0,1,2} {
            \foreach \y in {0,1,2} {
                \node[sitenode] at (\x,\y) {};
            }
        }
    }

    % ==================== (a) Merging two loops ====================
    % Panel label at upper left
    \node[font=\bfseries, anchor=north west] at (-0.3, 2.5) {(a)};

    % Before G - left plaquette with background
    \begin{scope}[shift={(0,0)}]
        \node[labelnode] at (1, -0.7) {$G$};
        % Background highlight for the loop
        \fill[bgblue, rounded corners=2pt] (-0.15,0.85) rectangle (1.15,2.15);
        \drawgrid
        % Loop edges (blue, thick)
        \draw[loopedge] (0,2) -- (1,2) -- (1,1) -- (0,1) -- cycle;
    \end{scope}

    % XOR symbol
    \node[opnode] at (3, 1.5) {$\oplus$};

    % Plaquette P with background
    \begin{scope}[shift={(4,0)}]
        \node[labelnode] at (1, -0.7) {$P$};
        % Background highlight for the plaquette
        \fill[bgorange, rounded corners=2pt] (0.85,0.85) rectangle (2.15,2.15);
        \drawgrid
        % Plaquette edges (orange, dashed)
        \draw[plaqedge] (1,2) -- (2,2) -- (2,1) -- (1,1) -- cycle;
    \end{scope}

    % Equals sign
    \node[opnode] at (7, 1.5) {$=$};

    % Result G'
    \begin{scope}[shift={(8,0)}]
        \node[labelnode] at (1, -0.7) {$G' = G \oplus P$};
        % Background highlight for merged loop
        \fill[bgblue, rounded corners=2pt] (-0.15,0.85) rectangle (2.15,2.15);
        \drawgrid
        % Merged loop (shared edge removed)
        \draw[loopedge] (0,2) -- (1,2) -- (2,2) -- (2,1) -- (1,1) -- (0,1) -- cycle;
    \end{scope}

    % ==================== (b) Splitting a loop ====================
    \begin{scope}[shift={(0,-4)}]
        % Panel label at upper left
        \node[font=\bfseries, anchor=north west] at (-0.3, 2.5) {(b)};

        % Before G - rectangle
        \begin{scope}[shift={(0,0)}]
            \node[labelnode] at (1, -0.7) {$G$};
            \fill[bgblue, rounded corners=2pt] (-0.15,0.85) rectangle (2.15,2.15);
            \drawgrid
            % Rectangle loop
            \draw[loopedge] (0,2) -- (1,2) -- (2,2) -- (2,1) -- (1,1) -- (0,1) -- cycle;
        \end{scope}

        % XOR symbol
        \node[opnode] at (3, 1.5) {$\oplus$};

        % Plaquette P (left one)
        \begin{scope}[shift={(4,0)}]
            \node[labelnode] at (1, -0.7) {$P$};
            \fill[bgorange, rounded corners=2pt] (-0.15,0.85) rectangle (1.15,2.15);
            \drawgrid
            % Left plaquette
            \draw[plaqedge] (0,2) -- (1,2) -- (1,1) -- (0,1) -- cycle;
        \end{scope}

        % Equals sign
        \node[opnode] at (7, 1.5) {$=$};

        % Result G' - just right plaquette
        \begin{scope}[shift={(8,0)}]
            \node[labelnode] at (1, -0.7) {$G' = G \oplus P$};
            \fill[bgblue, rounded corners=2pt] (0.85,0.85) rectangle (2.15,2.15);
            \drawgrid
            % Right plaquette only
            \draw[loopedge] (1,2) -- (2,2) -- (2,1) -- (1,1) -- cycle;
        \end{scope}
    \end{scope}

    % ==================== (c) Winding cycle move on torus ====================
    \begin{scope}[shift={(0,-8)}]
        % Panel label at upper left
        \node[font=\bfseries, anchor=north west] at (-0.3, 2.5) {(c)};

        % Before G - empty configuration
        \begin{scope}[shift={(0,0)}]
            \node[labelnode] at (1, -0.7) {$G = \emptyset$};
            \drawgrid
            % PBC indicators (curved arrows)
            \draw[-stealth, gridgray, thick, rounded corners] (2.2, 1.5) -- (2.5, 1.5) -- (2.5, 2.7) -- (-0.5, 2.7) -- (-0.5, 1.5) -- (-0.2, 1.5);
            \draw[-stealth, gridgray, thick, rounded corners] (1.5, 2.2) -- (1.5, 2.5) -- (2.7, 2.5) -- (2.7, -0.5) -- (1.5, -0.5) -- (1.5, -0.2);
        \end{scope}

        % XOR symbol
        \node[opnode] at (3, 1.5) {$\oplus$};

        % Winding cycle W_h
        \begin{scope}[shift={(4,0)}]
            \node[labelnode] at (1, -0.7) {$W_h$};
            \drawgrid
            % Horizontal winding cycle (middle row, wraps around)
            \draw[windedge] (0,1) -- (2,1);
            % Wrap indicators
            \draw[windedge] (2,1) -- (2.3,1);
            \draw[windedge] (-0.3,1) -- (0,1);
            % Curved arrow showing wrap
            \draw[-stealth, windgreen, thick, line width=1.5pt] (2.4, 1) arc (180:0:0.3) node[midway, above, font=\tiny] {};
            \draw[-stealth, windgreen, thick, line width=1.5pt] (-0.4, 1) arc (0:180:0.3);
        \end{scope}

        % Equals sign
        \node[opnode] at (7, 1.5) {$=$};

        % Result G' - winding configuration
        \begin{scope}[shift={(8,0)}]
            \node[labelnode] at (1, -0.7) {$G' = W_h$};
            \drawgrid
            % Winding loop (horizontal, middle row)
            \draw[loopedge] (0,1) -- (2,1);
            % Wrap indicators (dotted)
            \draw[loopblue, line width=2pt, dotted, line cap=round] (2,1) -- (2.4,1);
            \draw[loopblue, line width=2pt, dotted, line cap=round] (-0.4,1) -- (0,1);
            % Label for winding number
            \node[font=\footnotesize, text=loopblue] at (1, -1.3) {winding $(1,0)$};
        \end{scope}
    \end{scope}
\end{tikzpicture}
\caption{MCMC moves via symmetric difference ($\oplus$) on loop configurations. Blue solid lines show the current configuration $G$ and orange/green dashed lines show the cycle to flip. Shaded regions highlight the active loops. (a)~Merging loops. Flipping an adjacent plaquette merges it with the existing loop by removing the shared edge. (b)~Splitting a loop. Flipping a plaquette inside a larger loop splits off that portion. (c)~Winding cycle move on a torus with periodic boundary conditions. XOR with the horizontal winding cycle $W_h$ creates a winding loop.}
\label{fig:plaquette_flip}
\end{figure*}

For lattice systems with elementary plaquettes, we employ plaquette flip updates. Given the current configuration $G$, we select a random plaquette $P$ and propose $G' = G \oplus P$ (symmetric difference). The symmetric difference operation $\oplus$ toggles each edge. The edges present in exactly one of $G$ or $P$ appear in $G'$, while the edges present in both are removed. Since each plaquette is a cycle and the symmetric difference of two even-degree subgraphs yields another even-degree subgraph, this operation preserves the loop constraint. Flipping a plaquette adjacent to an existing loop merges them into a larger loop by removing the shared edge and flipping a plaquette inside a loop splits off that portion (Fig.~\ref{fig:plaquette_flip}). These moves allow the MCMC to efficiently explore the space of loop configurations by growing, shrinking, merging, or splitting loops.

The Metropolis acceptance probability for a plaquette flip is defined as 
\begin{equation}
    A(G \to G') = \min\left(1, \frac{\prod_{e \in G'} u_e}{\prod_{e \in G} u_e}\right).
    \label{eq:metropolis_acceptance}
\end{equation}
For uniform edge weights, this simplifies to $A(G \to G') = \min(1, u^{|G'| - |G|})$, where $|G'| - |G|$ is the change in the number of edges. This acceptance probability satisfies detailed balance with respect to the target distribution $\pi(G) \propto \prod_{e \in G} u_e$ (Supplemental Material~\cite{supplemental}).

For lattices with periodic boundary conditions (PBC), the cycle basis must include winding cycles that wrap around the torus. In an $L \times L$ torus, we include two winding cycles in addition to the $(L^2 - 1)$ independent plaquettes, the horizontal winding cycle ($W_h$) and the vertical winding cycle ($W_v$). All $L$ horizontal (vertical) edges in a single row (column) form a loop that wraps once around the $x$($y$)-direction. An XOR operation with these winding cycles generates loop configurations with non-zero winding numbers $(w_x, w_y)$, where $w_x$ counts how many times the loop winds around the $x$-direction and $w_y$ counts windings in the $y$-direction. Including winding cycles ensures that the sampler can explore all topological sectors of the system. A single winding move transforms a trivial configuration into one that wraps around the torus (Fig.~\ref{fig:plaquette_flip}c).

Single-plaquette flips can become inefficient for large systems at low temperatures, where typical loop configurations span many plaquettes. To improve mixing, we supplement single-cycle moves with multi-cycle updates that flip $k$ randomly chosen cycles simultaneously:
\begin{equation}
    G' = G \oplus P_1 \oplus P_2 \oplus \cdots \oplus P_k,
\end{equation}
where $k$ is drawn uniformly from $\{1, 2, \ldots, k_{\max}\}$ and each $P_i$ is a randomly selected cycle from the cycle basis. Since the symmetric difference is associative and commutative, and XOR operation of any number of even-degree subgraphs yields another even-degree subgraph, multi-cycle moves preserve the loop constraint. Any valid loop configuration can be reached from any other through a sequence of XOR operations on the basis of cycle and winding. This follows from the fact that the basis cycles span the entire cycle space over $\mathrm{GF}(2)$, the Galois field with two elements, where addition is XOR. On an $L \times L$ torus, the cycle space has dimension $L^2 + 1$, comprising $(L^2 - 1)$ independent plaquettes plus two winding cycles. Since every even-degree subgraph can be uniquely decomposed as an XOR combination of these bases, the MCMC is ergodic.

At low temperatures where $|u| \to 1$, the loop sum is dominated by large configurations. To ensure adequate sampling of the empty graph which is needed for partition function estimation, we employ umbrella sampling~\cite{torrie1977nonphysical} with bias potential as
\begin{equation}
    W(G) = \gamma \cdot |G| \cdot \omega,
\end{equation}
where $\gamma \in [0,1]$ is a tuning parameter, $|G|$ is the number of edges in configuration $G$, and $\omega$ is the bias strength per edge.

The choice of $\omega$ is critical for efficient sampling. A configuration $G$ with $n$ edges has weight approximately
\begin{equation}
    w(G) \approx |\bar{u}|^n,
\end{equation}
where $\bar{u}$ is the average edge weight. When $|\bar{u}| < 1$ which is typical for ferromagnetic Ising models where BP captures most correlations, configuration weights slowly decrease, causing the distribution to spread across many configurations and making the empty graph rarely sampled.

The key insight is that the bias should counteract the natural edge weight scaling. Setting $\omega = -\log|\bar{u}| = \log(1/|\bar{u}|)$, the bias factor per edge becomes
\begin{equation}
    e^{-\omega} = |\bar{u}|,
\end{equation}
and the biased weight for a configuration with $n$ edges is
\begin{equation}
    \pi_{\mathrm{biased}}(G) \propto |\bar{u}|^n \cdot e^{-\gamma n \omega} = |\bar{u}|^n \cdot |\bar{u}|^{\gamma n} = |\bar{u}|^{(1+\gamma)n}.
    \label{eq:biased_weight}
\end{equation}
The parameter $\gamma$ controls the effective decay rate. Therefore, $1/|\bar{u}|$ provides the natural unit for the bias strength.

When sampling with acceptance probability $\min(1, e^{\Delta \log w})$, the Markov chain converges to the stationary distribution $P(G) = u^{|G|}/Z_{\mathrm{loop}}$, where the normalization constant is precisely the partition function. The empty graph has weight $u^{0} = 1$, so its stationary probability is $P(\emptyset) = 1/Z_{\mathrm{loop}}$. By the ergodic theorem, we find that $n_{\mathrm{empty}}/n_{\mathrm{total}} = P(\emptyset) = 1/Z_{\mathrm{loop}}$. Inverting this relation gives the estimator $Z_{\mathrm{loop}} = n_{\mathrm{total}}/n_{\mathrm{empty}}$. This approach exploits the empty graph as a reference state with known weight, allowing us to infer the unknown normalization constant from sampling statistics alone.

The partition function with finite bias is given as
\begin{equation}
    Z_{\mathrm{loop}} = \frac{n_{\mathrm{total}}}{n_{\mathrm{empty}}} \times \langle e^{W(G)} \rangle_W,
    \label{eq:Z_estimator}
\end{equation}
where $n_{\mathrm{empty}}$ counts the samples with $G = \emptyset$ and the average is over the biased ensemble~\cite{shirts2008statistically,chodera2007use}.

The loop MCMC sampling procedure is summarized in Algorithm~1. The algorithm initializes from the empty graph and performs Metropolis updates using cycles from the precomputed cycle basis. Each proposed move $G' = G \oplus C$ toggles all edges in cycle $C$, with the log-weight change computed as
\begin{equation}
    \Delta \log w = \sum_{e \in C \setminus G} \log u_e - \sum_{e \in C \cap G} \log u_e,
\end{equation}
where the first sum is over edges added and the second over edges removed. With umbrella sampling, the acceptance probability becomes $\min(1, \exp(\Delta \log w - \Delta W))$, where $\Delta W = W(|G'|) - W(|G|)$ is the bias potential change.

\begin{figure}[t]
\centering
\begin{minipage}{0.95\columnwidth}
\hrule\vspace{0.5em}
\textbf{Algorithm 1:} Loop MCMC for partition function estimation
\vspace{0.3em}\hrule\vspace{0.5em}
\label{alg:loop_mcmc}
\begin{algorithmic}[1]
\Require Edge weights $\{u_e\}$, cycle basis $\mathcal{C}$, bias parameters $\gamma$, $\omega$
\Ensure Estimate of $Z_{\mathrm{loop}}$
\State $G \gets \emptyset$ \Comment{Initialize empty configuration}
\For{$i = 1$ to $n_{\mathrm{burn}}$} \Comment{Burn-in phase}
    \For{$j = 1$ to $n_{\mathrm{local}}$}
        \State $C \gets$ random cycle from $\mathcal{C}$
        \State $G' \gets G \oplus C$
        \State $\Delta \log w \gets \sum_{e \in C \setminus G} \log u_e - \sum_{e \in C \cap G} \log u_e$
        \State $\Delta W \gets \gamma \omega (|G'| - |G|)$
        \If{$\log(\mathrm{rand}()) < \Delta \log w - \Delta W$}
            \State $G \gets G'$
        \EndIf
    \EndFor
\EndFor
\State $n_{\mathrm{empty}} \gets 0$, $S \gets 0$
\For{$i = 1$ to $n_{\mathrm{sweeps}}$} \Comment{Sampling phase}
    \For{$j = 1$ to $n_{\mathrm{local}}$}
        \State Perform Metropolis update (lines 4--10)
    \EndFor
    \State $S \gets S + \exp(\gamma \omega |G|)$ \Comment{Accumulate reweighting}
    \If{$G = \emptyset$}
        \State $n_{\mathrm{empty}} \gets n_{\mathrm{empty}} + 1$
    \EndIf
\EndFor
\State \Return $Z_{\mathrm{loop}} = (n_{\mathrm{sweeps}} / n_{\mathrm{empty}}) \times (S / n_{\mathrm{sweeps}})$
\end{algorithmic}
\vspace{0.3em}\hrule
\end{minipage}
\end{figure}

Thermodynamic quantities can be computed by combining automatic differentiation with MCMC-estimated loop statistics. Since $Z = Z_{\mathrm{BP}} \cdot Z_{\mathrm{loop}}$, the free energy separates as $F = F_{\mathrm{BP}} + F_{\mathrm{loop}}$, and thermodynamic derivatives related to $Z_{\mathrm{BP}}$ can be obtained using automatic differentiation. However, we cannot auto-differentiate $Z_{\mathrm{loop}}$ through a Monte Carlo estimate. 

For uniform edge weights $u(\beta)$, the loop partition function is $Z_{\mathrm{loop}} = \sum_{G \in \mathcal{L}} u^{|G|}$. The first and second derivatives with respect to $\beta$ involve the mean and variance of the edge count $|G|$ under the loop distribution as
\begin{align}
    \frac{\partial \log Z_{\mathrm{loop}}}{\partial \beta} &= \frac{1}{u}\frac{du}{d\beta} \langle |G| \rangle_{\mathrm{loop}}, \label{eq:dlogZ_loop} \\
    \frac{\partial^2 \log Z_{\mathrm{loop}}}{\partial \beta^2} &= \left(\frac{d^2 u/d\beta^2}{u} - \frac{(du/d\beta)^2}{u^2}\right) \langle |G| \rangle_{\mathrm{loop}} \notag \\
    &\quad + \frac{(du/d\beta)^2}{u^2} \mathrm{Var}(|G|)_{\mathrm{loop}}, \label{eq:d2logZ_loop}
\end{align}
where $\langle |G| \rangle_{\mathrm{loop}}$ and $\mathrm{Var}(|G|)_{\mathrm{loop}} = \langle |G|^2 \rangle - \langle |G| \rangle^2$ are estimated from MCMC samples using the stored reweighting factors. As shown in Eq.~(\ref{eq:d2logZ_loop}), computing the second derivative $\partial^2 \log Z_{\mathrm{loop}} / \partial \beta^2$ requires the variance of the edge count $\mathrm{Var}(|G|)$ under the unbiased loop distribution. Since MCMC samples are drawn from the biased ensemble, this variance must be estimated by reweighting the importance sampling.
% [Eq.~(\ref{eq:reweight_mean})]. 

The loop contribution to energy and specific heat depends on the first and second moments of the edge count $|G|$ under the \emph{unbiased} loop distribution [Eqs.~(\ref{eq:dlogZ_loop})--(\ref{eq:d2logZ_loop})]. However, MCMC samples are drawn from the \emph{biased} distribution with umbrella potential $W(G) = \gamma \omega |G|$. To recover unbiased expectations, the importance sampling is reweighted as
\begin{equation}
    \langle |G| \rangle_{\mathrm{loop}} = \frac{\langle |G| \cdot e^{W(G)} \rangle_W}{\langle e^{W(G)} \rangle_W},
    \label{eq:reweight_mean}
\end{equation}
and similarly for $\langle |G|^2 \rangle$. The reweighting factor $e^{W(G)}$ can vary by many orders of magnitude, making the effective sample size much smaller than the actual number of samples~\cite{kong1992note}.

For large systems, the umbrella sampling bias introduces exponentially large reweighting factors, causing the effective sample size to collapse and the variance estimate to become unreliable. This signal-to-noise degradation currently limits accurate computation of energy and specific heat for small systems. 

\begin{figure*}[t]
\centering
\includegraphics[width=\textwidth]{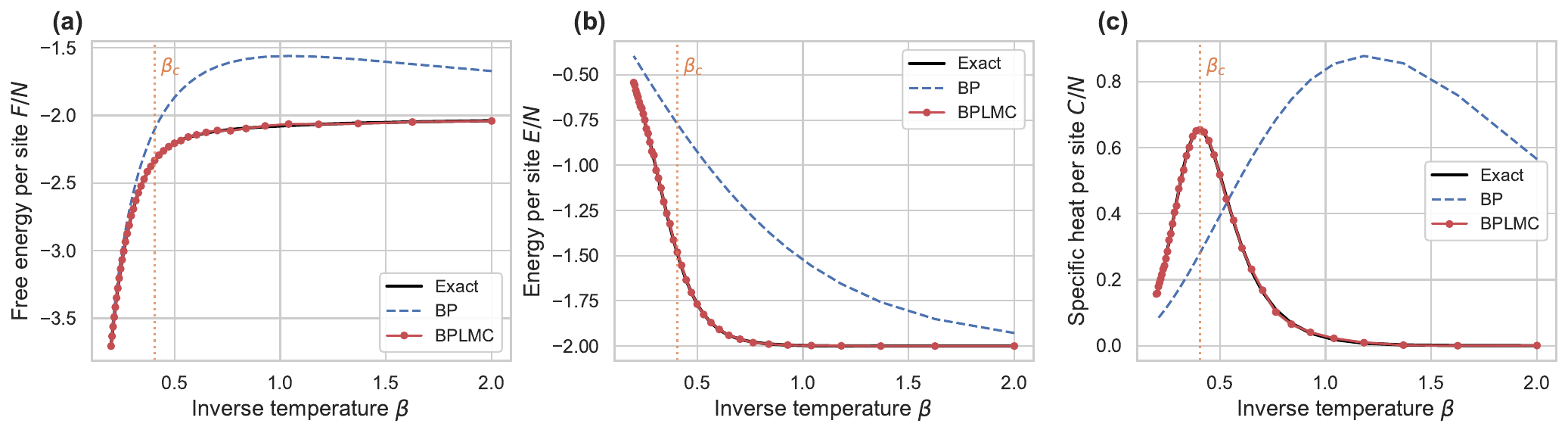}
\caption{Comparison of exact enumeration (black solid), BP (blue dashed), and BPLMC (red line) for the ($3 \times 3$) ferromagnetic Ising model with periodic boundary conditions. (a) Free energy per site $F/N$. (b) Energy per site $E/N$. (c) Specific heat per site $C/N$. All quantities are plotted versus inverse temperature $\beta$. BPLMC matches the exact solution across all temperatures, while BP shows systematic errors that grow at low temperature (high $\beta$).}
\label{fig:3x3_comparison}
\end{figure*}

%All calculations were performed using PyTorch~\cite{paszke2019pytorch} and the implementation is available as the \texttt{knots} package at \url{https://github.com/myung-group/knots}. 
% Results

\section{Results}
\label{sec:results}

\subsection{Benchmark: 3$\times$3 lattice with exact comparison}
\label{subsec:3x3}
We first validate our BPLMC on a $3 \times 3$ square lattice with PBC, where exact enumeration of all $2^9 = 512$ spin configurations is feasible. This small system serves as an ideal test case because exact results are available for rigorous validation, and the system is large enough to exhibit non-trivial loop topology. The $3 \times 3$ lattice with PBC has $N = 9$ sites and $|E| = 18$ edges. In the loop representation, configurations are even-degree subgraphs of this lattice. The cycle space has dimension $\dim(\mathcal{C}) = |E| - |V| + 1 = 18 - 9 + 1 = 10$, decomposing into $(L^2 - 1) = 8$ independent plaquettes and 2 topologically non-trivial winding cycles (horizontal and vertical).

We find several key observations emerge at different temperature regimes in the $3 \times 3$ ferromagnetic Ising model by comparing exact, BP, and BPLMC results (Fig.~\ref{fig:3x3_comparison} and Table~S1 in the Supplemental Material~\cite{supplemental}). At high temperatures (small $\beta$), the loop weight $u = \tanh(\beta J)$ is small, so configurations with edges are exponentially suppressed. The empty graph dominates, giving $Z_{\mathrm{loop}} \approx 1$. In this regime, BP is already reasonably accurate for the free energy (Fig.~\ref{fig:3x3_comparison}a), and BPLMC provides consistent improvements. Near the critical temperature ($\beta_c \approx 0.41$, estimated from the specific heat divergence), loop correlations become significant as correlations extend throughout the system. The BP error for the free energy grows large, while BPLMC maintains high accuracy. At low temperatures (high $\beta$), the edge weight approaches unity, which makes large loop configurations important. Here, the BP free energy error becomes substantial, while BPLMC continues to match the exact solution.

The energy per site (Fig.~\ref{fig:3x3_comparison}b) shows similar trends. BP systematically overestimates the energy at all temperatures, while BPLMC matches the exact solution. The energy error in BP grows as temperature decreases because the Bethe approximation increasingly fails to capture the strong spin correlations that develop in the ordered phase. In the specific heat comparison, BP shows no sign of specific heat divergence (only a spurious broad peak at high $\beta$) instead of the correct peak near the finite-size critical point $\beta_c \approx 0.41$, because it ignores loop correlations that dominate the critical fluctuations (Fig.~\ref{fig:3x3_comparison}c). The vertical dotted orange line marks $\beta_c$, determined from the heat capacity peak, which is shifted from the thermodynamic limit value ($\beta_c \approx 0.44$) due to finite size effects. BPLMC (red with shaded error bands) accurately tracks the exact solution across all temperatures, correctly capturing both the peak position and magnitude.

\begin{figure*}[t]
\centering
\includegraphics[width=0.85\textwidth]{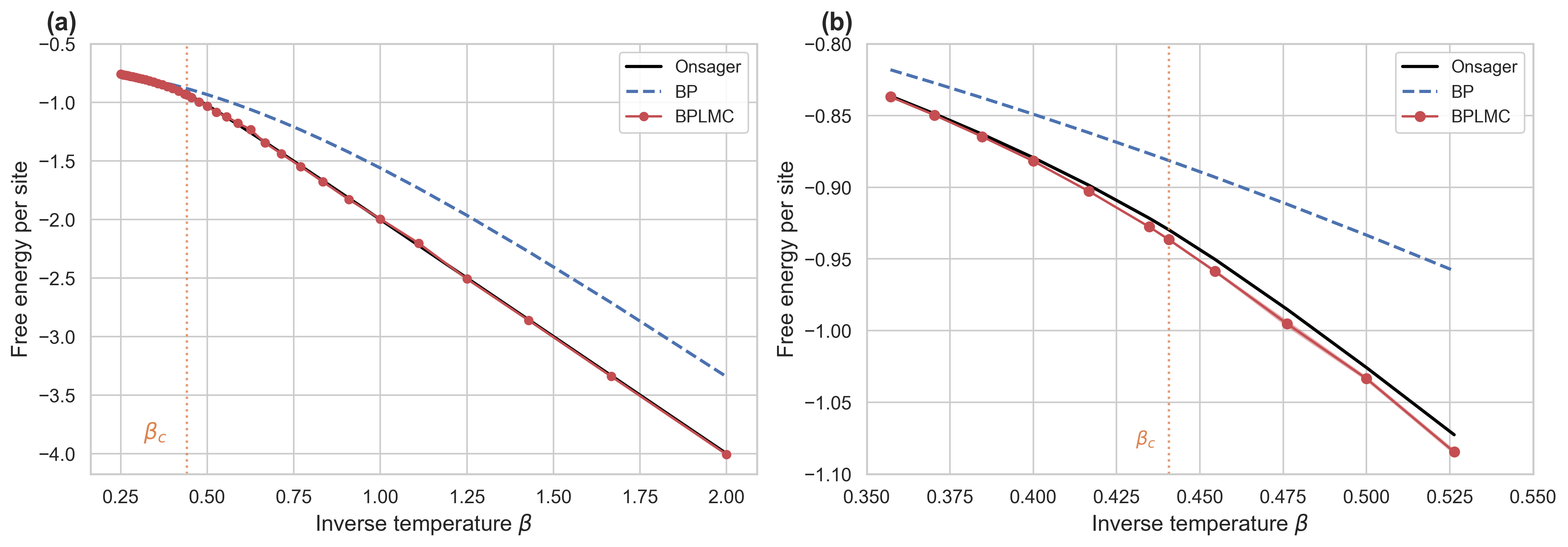}
\caption{Free energy per site for the ($10 \times 10$) ferromagnetic Ising model with periodic boundary conditions. (a) Full inverse temperature range showing BP (blue dashed), BPLMC (red), and Onsager exact solution (black solid). (b) Magnified view around the critical temperature $\beta_c \approx 0.44$ (orange dotted line). BPLMC accurately tracks the exact solution across all temperatures, while BP shows systematic deviations that persist even at low temperature (high $\beta$). The continued deviation of BP at low temperature arises because we initialize messages uniformly, which corresponds to the high-temperature fixed point. For $\beta > \beta_{\mathrm{BP}} = \ln(2)/2 \approx 0.347$, BP admits multiple fixed points~\cite{midha2025beyond}, and the correct low-temperature fixed point requires symmetry-broken initialization to capture the ferromagnetic ordering.}
\label{fig:10x10_free_energy}
\end{figure*}

BP (blue dashed line) predicts a spurious broad peak at $\beta \approx 1$, far from the true critical point. This artifact arises fundamentally because the BP solution used here corresponds to the high-temperature (paramagnetic) fixed point, which we maintain throughout all temperatures. In the exact theory, below the critical temperature, the system transitions to an ordered phase with broken symmetry and long-range correlations. However, by enforcing the paramagnetic fixed point, BP continues to describe a disordered state even in the low-temperature regime where this solution is no longer physical. This mismatch between the assumed paramagnetic state and the true ordered ground state leads to systematic errors that grow with decreasing temperature. The spurious heat capacity peak also reflects BP's incorrect description of energy fluctuations in this regime. Since $C = \beta^2 (\langle E^2 \rangle - \langle E \rangle^2)$, errors in variance are amplified, and the failure of the paramagnetic fixed point to capture the onset of order produces an artificial variance peak. The two-dimensional ferromagnetic Ising model has exactly one phase transition at $\beta_c \approx 0.44$~\cite{onsager1944crystal}, and BPLMC, which computes the exact partition function, correctly shows no peak at $\beta \approx 1$. While Midha and Zhang~\cite{midha2025beyond} showed that BP fixed points undergo a bifurcation at $\beta_{\mathrm{BP}} = \ln(2)/2 \approx 0.347$, the spurious BP peak occurs at a different temperature, further confirming it is an artifact of using the wrong fixed point rather than a reflection of any BP critical behavior.

\begin{figure*}[t]
\centering
\includegraphics[width=\textwidth]{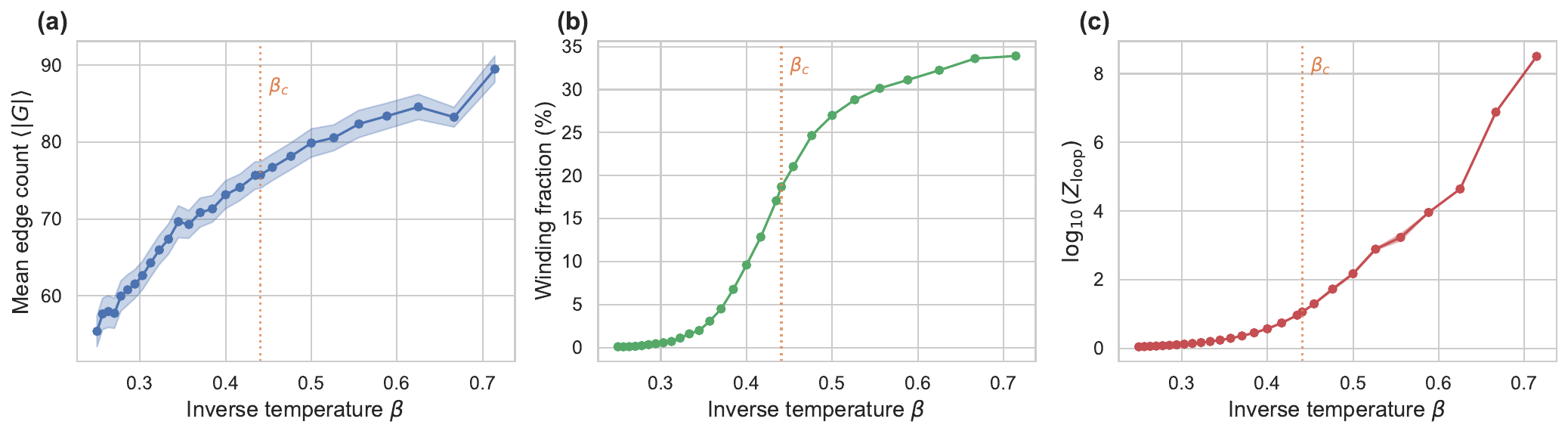}
\caption{Temperature dependence of loop configuration statistics for the $10 \times 10$ ferromagnetic Ising model. (a) Mean edge count $\langle |G| \rangle$ increases with inverse temperature $\beta$, reflecting the growing statistical weight of configurations with more edges as temperature decreases. (b) Winding fraction, the percentage of loop configurations containing topologically non-trivial cycles that wrap around the torus. The winding fraction increases sharply near the critical temperature $\beta_c \approx 0.44$ (vertical dotted line), indicating a transition from locally fluctuating plaquettes to system-spanning correlations. (c) Logarithm of the loop partition function $Z_{\mathrm{loop}} = Z/Z_{\mathrm{BP}}$, which quantifies the correction factor from BP to the exact partition function. Near $\beta_c$, $\log_{10} Z_{\mathrm{loop}}$ grows rapidly, indicating that loop corrections become increasingly important as the system approaches criticality.}
\label{fig:loop_statistics}
\end{figure*}

The MCMC estimator exhibits the expected $1/\sqrt{N}$ error scaling with the number of samples $N$, as verified at the critical temperature $\beta_c \approx 0.4407$ where sampling is challenging (see Supplemental Material~\cite{supplemental}). At $10^5$ sweeps, statistical errors reach $\sim 0.003$ for $F/N$, $\sim 0.005$ for $E/N$, and $\sim 0.004$ for $C/N$. The agreement between statistical error estimates (from block averaging for $E$ and $C$, jackknife resampling for $F$) and actual errors versus exact values confirms that our estimators are unbiased~\cite{flyvbjerg1989error,QUENOUILLE1956notes}.

\subsection{Benchmark: $10 \times 10$ lattice versus Onsager solution}
\label{subsec:10x10}

For larger systems where exact enumeration is infeasible, we compare against the Onsager exact solution~\cite{onsager1944crystal} for the two-dimensional Ising model in the thermodynamic limit. The Onsager free energy per site is
\begin{align}
    f_{\mathrm{Ons}} &= -\frac{1}{\beta}\biggl[\ln(2\cosh 2K) \notag \\
    &\quad + \frac{1}{\pi} \int_0^{\pi/2} \ln\frac{1 + \sqrt{1 - \kappa^2 \sin^2\theta}}{2} \, d\theta\biggr],
    \label{eq:onsager}
\end{align}
where $K = \beta J$ and $\kappa = 2\sinh(2K)/\cosh^2(2K)$.

Figure~\ref{fig:10x10_free_energy} presents the free energy comparison for a $10 \times 10$ lattice. The BP approximation shows significant deviations from the Onsager solution, particularly around the critical region. Notably, the BP error persists even at low temperatures (high $\beta$), in contrast to what one might expect from a mean-field theory that should become accurate deep in the ordered phase. As we discussed in Sec.~\ref{subsec:3x3}, this behavior arises because our BP implementation uses uniform message initialization, which corresponds to the high-temperature (paramagnetic) fixed point. Our uniform initialization causes the message-passing iteration to converge to the paramagnetic fixed point even at low temperatures, which explains the persistent BP error. However, the BPLMC method successfully corrects these errors by properly sampling loop configurations, achieving excellent agreement with the Onsager solution at all temperatures.

Table~S2 in the Supplemental Material~\cite{supplemental} presents detailed numerical results, comparing BP and BPLMC against the Onsager solution. The results reveal important characteristics of the method. First, BPLMC consistently improves BP from high temperature to low temperature. Second, finite-size effects are visible as the $10 \times 10$ lattice differs from the thermodynamic limit, so both BP and BPLMC show residual errors. BPLMC computes the correct finite-size partition function, which differs slightly from Onsager's infinite-system result. Third, sampling becomes more challenging for larger systems, which requires umbrella sampling for reliable estimation.

\subsection{Analysis of loop configuration statistics}
\label{subsec:loop_stats}

To understand why BPLMC succeeds, we analyze the statistics of sampled loop configurations. Figure~\ref{fig:loop_statistics} shows the temperature dependence of three key quantities for the ($10 \times 10$) system, including the mean edge count, the winding fraction, and the loop partition function.

Figure~\ref{fig:loop_statistics}a shows the mean edge count $\langle |G| \rangle$ as a function of inverse temperature $\beta$. At high temperature ($\beta \lesssim 0.3$), the mean edge count is approximately $60$--$70$ edges out of the 180 total lattice edges. As temperature decreases toward the critical point $\beta_c \approx 0.44$ (marked by the vertical dotted line), $\langle |G| \rangle$ increases monotonically, reaching approximately $90$ edges at $\beta = 0.7$. This systematic increase reflects the growing edge weight $u = \tanh(\beta J)$. Since a configuration with $n$ edges has weight $u^n$, smaller $u$ (high temperature) exponentially suppresses configurations with many edges, while as $u$ approaches unity (low temperature), this suppression weakens and the combinatorially larger number of configurations with many edges shifts the distribution toward higher $\langle |G| \rangle$.

Figure~\ref{fig:loop_statistics}b reveals the winding fraction which is the proportion of loop configurations containing at least one winding loop that wraps around the cell. This quantity exhibits striking temperature dependence. At high temperature ($\beta \lesssim 0.35$), the winding fraction is negligible, indicating that virtually all configurations consist entirely of small loops, namely plaquettes. Near $\beta_c$, the winding fraction increases sharply, signaling the emergence of system-spanning correlations. This sharp onset provides a signature of the phase transition. As the correlation length grows to match the system size, winding loops that encircle the entire cell become statistically significant. The winding fraction continues to grow at lower temperatures, reaching substantial values in the ordered phase.

Figure~\ref{fig:loop_statistics}c shows $\log_{10} Z_{\mathrm{loop}}$, the logarithm of the loop correction factor $Z_{\mathrm{loop}} = Z / Z_{\mathrm{BP}}$. This quantity directly measures the magnitude of the correction from BP to the exact partition function. At high temperature, $Z_{\mathrm{loop}} \approx 1$ ($\log_{10} Z_{\mathrm{loop}} \approx 0$), confirming that BP is accurate when loop correlations are weak. As $\beta$ increases toward $\beta_c$, $\log_{10} Z_{\mathrm{loop}}$ grows rapidly, reaching values exceeding unity. This exponential growth in $Z_{\mathrm{loop}}$ explains why BP errors become substantial near criticality. The loop contributions that BP ignores carry an increasingly large fraction of the partition function weight.

\begin{figure*}[t]
\centering
\includegraphics[width=\textwidth]{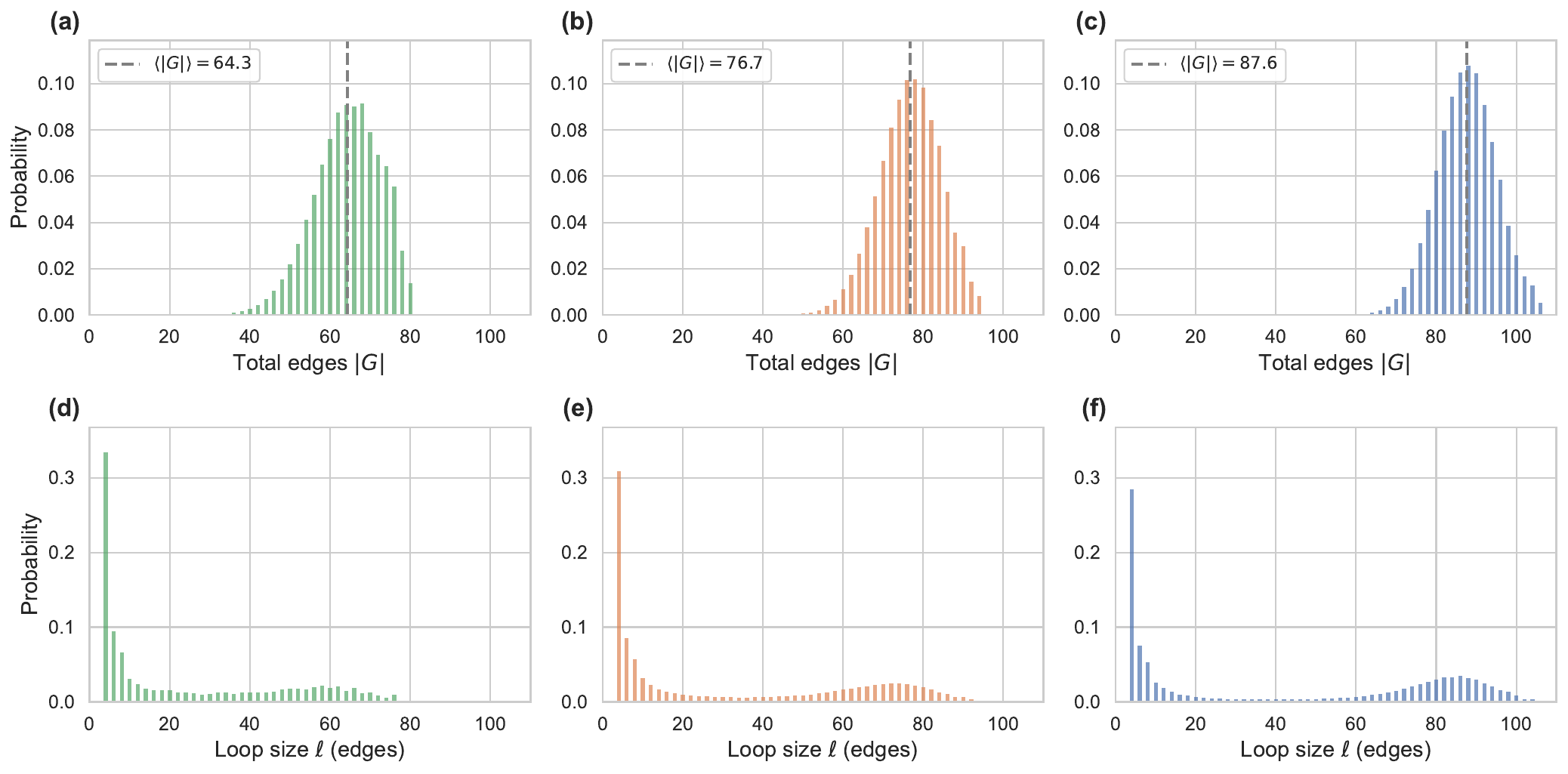}
\caption{Distribution of loop configurations for the ($10 \times 10$) ferromagnetic Ising model at three representative temperatures. Distribution of total edge count $|G|$ per configuration at (a) high temperature $\beta = 0.3125$, (b) near critical temperature $\beta = 0.4545 \approx \beta_c$, and (c) low temperature $\beta = 0.7143$. The grey dashed line indicates the reweighted mean $\langle |G| \rangle$. Distribution of individual loop sizes $\ell$ (number of edges per connected component) at the same temperatures. At high temperature (d), small contractible loops of size 4 (plaquettes) dominate, with a secondary peak at larger sizes corresponding to system-spanning winding loops. As temperature decreases (e, f), the winding loop peak shifts to larger sizes and becomes more prominent, reflecting the increasing importance of winding configurations in the ordered phase.}
\label{fig:histograms}
\end{figure*}

Figure~\ref{fig:histograms} provides more detailed insights into the loop configuration distributions at three representative temperatures. At high temperature ($\beta = 0.3125$, Fig.~\ref{fig:histograms}a), the distribution is broad with $\langle |G| \rangle \approx 64$ edges. Near criticality ($\beta \approx \beta_c$, Fig.~\ref{fig:histograms}b), the mean shifts to $\langle |G| \rangle \approx 77$ edges, while at low temperature ($\beta = 0.7143$, Fig.~\ref{fig:histograms}c), configurations with $\langle |G| \rangle \approx 88$ edges dominate.

The distribution of individual loop sizes $\ell$ which are the number of edges in each connected component reveals the internal structure of loop configurations. At all temperatures, the distribution exhibits a bimodal structure. The first peak at $\ell = 4$ corresponds to elementary plaquettes, which are the smallest contractible loops on the square lattice. The second peak at larger $\ell$ corresponds to winding loops. At high temperature (Fig.~\ref{fig:histograms}d), winding loops are relatively rare and have sizes around $\ell \approx 60$--$70$ edges. As temperature decreases (Fig.~\ref{fig:histograms}e and f), the winding loop peak shifts to larger sizes ($\ell \approx 80$--$100$ edges) and becomes more prominent relative to the plaquette peak. This bimodal structure has important physical implications. The small plaquettes represent local spin fluctuations. However, the large winding loops represent global correlations that span the entire system. By explicitly sampling all loop configurations, BPLMC correctly accounts for both local and global correlations.

% Discussion
\section{Discussion}
\label{sec:discussion}

We have introduced BPLMC, a hybrid method combining belief propagation with Monte Carlo sampling of loop corrections. The approach achieves accurate results, subject only to statistical sampling error. It includes all loop orders including long-range correlations without convergence concerns. Similarly, the PEPS cluster updates~\cite{lubasch2014unifying,lubasch2014algorithms} showed that increasing the cluster size $\delta$ systematically improves the environment approximation from BP to exact contraction, with the required $\delta$ scaling with the correlation length. This result is consistent with our observation that loop corrections become critical near the phase transition where correlations are long-ranged.
For systems where analytical methods fail, MCMC may be one of the viable paths to accurate results.

%A crucial technical contribution is the umbrella sampling scheme that maintains adequate empty-graph sampling at low temperatures. Without bias, the empty graph becomes exponentially rare as $\beta \to \infty$, making partition function estimation difficult. Our bias parameter $\omega$ ensures that even when $u \to 1$ (and $-\log u \to 0$), the bias remains effective. This enables BPLMC to work across the full temperature range.

Our $10 \times 10$ 2D ferromagnetic Ising model benchmarks reveal that the loop configurations sampled by our MCMC algorithm provide physical insight into the system's correlations. On a finite lattice of linear size $L$ with periodic boundary conditions, loop configurations fall into two topologically distinct classes: (1) local plaquette loops and (2) winding loops that wrap around the unit cell. The prevalence of winding configurations is connected to the correlation length $\xi$. When $\xi \ll L$, winding loops are suppressed, while near the critical point where $\xi \sim L$, winding loops contribute significantly.

When $\xi \ll L$ (high temperature, disordered phase), correlations are local and winding loops are suppressed by their extensive edge count. At the critical point where $\xi \sim L$, scale-invariant fluctuations allow winding loops to contribute significantly~\cite{pollock1987path}.
When $\xi \gg L$ (low temperature, ordered phase), the dominant loop corrections are non-local (winding), precisely the correlations that local methods like BP cannot capture. Our MCMC approach naturally samples these configurations, providing corrections that become important near $\beta_c$. 

The current implementation has three limitations that suggest directions for future work. First, sampling efficiency degrades for large systems at low temperatures, where mean loop sizes grow and single-plaquette moves become insufficient. We have implemented multi-plaquette moves and parallel tempering as partial solutions, but more sophisticated algorithms such as worm updates~\cite{prokofev2001worm} or cluster moves~\cite{wolff1989collective} may be needed for very large systems. Second, systems like frustrated antiferromagnets will introduce a sign problem in the edge basis. Third, while our formulation assumes symmetric edge potentials, the BPLMC framework can be extended to asymmetric potentials and multi-state variables through generalized loop expansions.

More broadly, BPLMC offers a general paradigm for tensor network contraction. One can use BP as a tractable reference and stochastically sample loop corrections to systematically recover the exact result. Although we have demonstrated the method on the 2D Ising model as a benchmark, the framework can be extended to arbitrary tensor networks that represent 2D and 3D quantum lattice models as well as ab initio molecular systems.

%mapped to a pairwise MRF with symmetric potentials, including partition functions of vertex models, quantum circuit amplitudes with real positive weights, and probabilistic graphical models arising in machine learning. This perspective connects to importance sampling in Monte Carlo~\cite{hammersley1964monte}, path-integral Monte Carlo~\cite{pollock1987path}, and diagrammatic Monte Carlo in many-body physics~\cite{prokofev1998polaron,vanhoucke2010diag,gull2011ctqmc}.

% Acknowledgments
\vspace{10pt}
% Acknowledgments
\begin{acknowledgments}
CWM acknowledges the support provided by the National Research Foundation of Korea (NRF) grants funded by the Korean government (MSIT) (Grant No.~RS-2023-00283929, RS-2022-NR072058). SYW and CWM acknowledge the support provided by the NRF grants funded by the Korean government (MSIT) (Grant No.~RS-2024-00407680). DCY acknowledges the support provided by the NRF grant funded by the Korean government (MSIT) (Grant No.~RS-2023-00250313). This research was also supported by `Quantum Information Science R\&D Ecosystem Creation' through the NRF funded by the Korean government (MSIT) (No. 2020M3H3A1110365). The authors are grateful for the computational resources at the Korea Institute of Science and Technology Information~(KISTI) with the \textsc{Nurion} cluster (KSC-2025-CRE-0286, KSC-2025-CRE-0316, KSC-2025-CRE-0125, KSC-2025-CRE-0122 ). Computational work for this research was also partially performed on the \textsc{Olaf} cluster supported by IBS Research Solution Center and on the GPU cluster supported by NIPA.
\end{acknowledgments}

% References
% \input{sections/references}
\bibliography{references}

\end{document}

\title{Supplemental Material for ``Stochastic Tensor Network Contraction Beyond Belief Propagation''}

\author{Gi Beom Sim}
\thanks{Contributed equally to this work}
\affiliation{Department of Energy Science, Sungkyunkwan University, Seobu-ro 2066, Suwon, 16419, Korea}

\author{Tae Hyeon Park}
\thanks{Contributed equally to this work}
\affiliation{Department of Energy Science, Sungkyunkwan University, Seobu-ro 2066, Suwon, 16419, Korea}

\author{Kwang S. Kim}
\affiliation{Department of Chemistry, Ulsan National Institute of Science and Technology, 50 UNIST-gil, Ulsan 44919, Republic of Korea}

\author{Yanmei Zang}
\affiliation{Department of Energy Science, Sungkyunkwan University, Seobu-ro 2066, Suwon, 16419, Korea}

\author{Xiaorong Zou}
\affiliation{Department of Energy Science, Sungkyunkwan University, Seobu-ro 2066, Suwon, 16419, Korea}

\author{Hye Jung Kim}
\email{hjkim75@skku.edu}
\affiliation{Department of Energy Science, Sungkyunkwan University, Seobu-ro 2066, Suwon, 16419, Korea}

\author{D. ChangMo Yang}
\email{dcyang@skku.edu}
\affiliation{Department of Energy Science, Sungkyunkwan University, Seobu-ro 2066, Suwon, 16419, Korea}

\author{Soohaeng Yoo Willow}
\email{sy7willow@gmail.com}
\affiliation{Department of Energy Science, Sungkyunkwan University, Seobu-ro 2066, Suwon, 16419, Korea}

\author{Chang Woo Myung}
\email{cwmyung@skku.edu}
\affiliation{Department of Energy Science, Sungkyunkwan University, Seobu-ro 2066, Suwon, 16419, Korea}
\affiliation{Department of Quantum Information Engineering, Sungkyunkwan University, Seobu-ro 2066, Suwon, 16419, Korea}
\affiliation{Department of Energy, Sungkyunkwan University, Seobu-ro 2066, Suwon, 16419, Korea}

\maketitle

\setcounter{equation}{0}
\setcounter{figure}{0}
\setcounter{table}{0}
\renewcommand{\theequation}{S\arabic{equation}}
\renewcommand{\thefigure}{S\arabic{figure}}
\renewcommand{\thetable}{S\arabic{table}}

%======================================================================
\section{Detailed Balance for Metropolis-Hastings Sampling}
\label{sec:detailed_balance}

We derive the Metropolis-Hastings acceptance probability for plaquette flip moves and verify that it satisfies detailed balance. We aim to sample loop configurations $G$ from the distribution $\pi(G) = w(G)/Z_{\mathrm{loop}}$, where $w(G) = \prod_{e \in G} u_e$ and $Z_{\mathrm{loop}} = \sum_{G \in \mathcal{L}} w(G)$ is the loop partition function. For a Markov chain to have $\pi$ as its stationary distribution, the transition probabilities $T(G \to G')$ must satisfy detailed balance: $\pi(G) \cdot T(G \to G') = \pi(G') \cdot T(G' \to G)$.

The transition probability factors into proposal and acceptance: $T(G \to G') = q(G \to G') \cdot A(G \to G')$. For plaquette flip moves, we select a plaquette $P$ uniformly at random and propose $G' = G \oplus P$ (symmetric difference). Since both forward and reverse moves select from the same set of plaquettes with equal probability, $q(G \to G') = q(G' \to G) = 1/n_{\mathrm{plaquettes}}$. The Metropolis choice that satisfies detailed balance is:
\begin{equation}
    A(G \to G') = \min\left(1, \frac{w(G')}{w(G)}\right) = \min\left(1, \frac{\prod_{e \in G'} u_e}{\prod_{e \in G} u_e}\right).
    \label{eq:acceptance}
\end{equation}

We verify this property by considering two cases. First, when $w(G') \geq w(G)$, the acceptance probabilities are $A(G \to G') = 1$ and $A(G' \to G) = w(G)/w(G')$. In this case,  the left-hand side is $\pi(G) \cdot q \cdot 1 = w(G) q / Z_{\mathrm{loop}}$, while the right-hand side is $\pi(G') \cdot q \cdot w(G)/w(G') = w(G) q / Z_{\mathrm{loop}}$. Therefore, the two sides are equal.

Second, when $w(G') < w(G)$, the acceptance probabilities becomes $A(G \to G') = w(G')/w(G)$ and $A(G' \to G) = 1$. In this case, both the left-hand side and right-hand side reduce to $w(G') q / Z_{\mathrm{loop}}$. Hence, two sides are equal.
Eventually, detailed balance is satisfied in both cases.

When all edges have the same weight $u_e = u$, the acceptance probability simplifies to $A(G \to G') = \min(1, u^{|G'| - |G|})$. For ferromagnetic systems where $0 < u < 1$, adding edges ($|G'| > |G|$) has acceptance probability $u^{|G'|-|G|} < 1$. Removing edges ($|G'| < |G|$) is always accepted, and moves with the same edge count ($|G'| = |G|$) are always accepted. 
%======================================================================

\section{Benchmark Data Table for the $3 \times 3$ ferromagnetic Ising model}
\label{sec:tables}

\begin{table}[h]
\caption{Free energy per site for the $3 \times 3$ ferromagnetic Ising model with periodic boundary conditions. Comparison of exact enumeration, belief propagation (BP), and BPLMC across inverse temperatures $\beta$. The critical point is $\beta_c \approx 0.41$.}
\label{tab:3x3}
\begin{ruledtabular}
\begin{tabular}{cccccc}
$\beta$ & $F_{\mathrm{exact}}/N$ & $F_{\mathrm{BP}}/N$ & $F_{\mathrm{BPLMC}}/N$ & BP err & BPLMC err \\
\hline
0.20 & $-3.705$ & $-3.664$ & $-3.706$ & $4.1 \times 10^{-2}$ & $3 \times 10^{-4}$ \\
0.23 & $-3.346$ & $-3.290$ & $-3.349$ & $5.6 \times 10^{-2}$ & $3 \times 10^{-3}$ \\
0.27 & $-2.937$ & $-2.851$ & $-2.935$ & $8.6 \times 10^{-2}$ & $2 \times 10^{-3}$ \\
0.33 & $-2.574$ & $-2.429$ & $-2.573$ & $1.4 \times 10^{-1}$ & $2 \times 10^{-3}$ \\
0.39 & $-2.377$ & $-2.165$ & $-2.377$ & $2.1 \times 10^{-1}$ & $1 \times 10^{-4}$ \\
0.45 & $-2.263$ & $-1.980$ & $-2.262$ & $2.8 \times 10^{-1}$ & $1 \times 10^{-3}$ \\
0.53 & $-2.182$ & $-1.814$ & $-2.186$ & $3.7 \times 10^{-1}$ & $4 \times 10^{-3}$ \\
0.70 & $-2.116$ & $-1.638$ & $-2.111$ & $4.8 \times 10^{-1}$ & $5 \times 10^{-3}$ \\
1.04 & $-2.074$ & $-1.560$ & $-2.063$ & $5.1 \times 10^{-1}$ & $1.1 \times 10^{-2}$ \\
2.00 & $-2.039$ & $-1.672$ & $-2.041$ & $3.7 \times 10^{-1}$ & $2 \times 10^{-3}$ \\
\end{tabular}
\end{ruledtabular}
\end{table}

\section{MCMC Convergence and Error Scaling}
\label{sec:convergence}

\begin{figure}[h]
\centering
\includegraphics[width=1.0\textwidth]{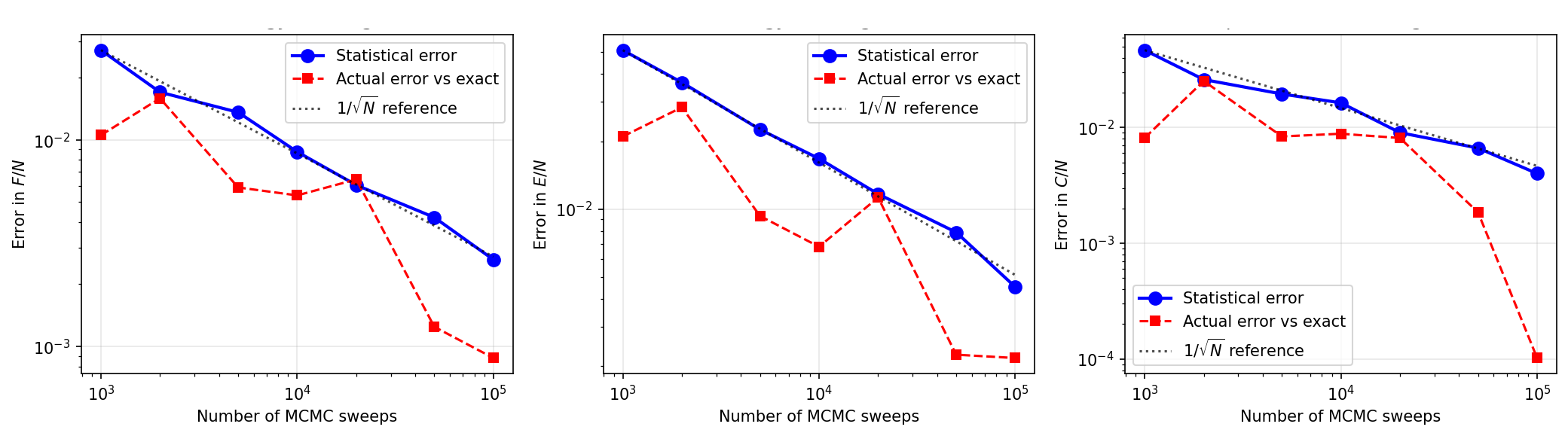}
\caption{Convergence of MCMC errors of free energy per site (left), energy per site (middle), and heat capacity (right) with number of sweeps for the $3 \times 3$ Ising model at $\beta = \beta_c$. Blue circles denote statistical errors from block averaging ($E$, $C$) or jackknife ($F$). Red squares denote actual errors vs exact values. Dotted line indicates $1/\sqrt{N}$ reference. Scaling exponents are approximately $-0.5$ for all quantities, confirming that the MCMC estimators are unbiased.}
\label{fig:convergence}
\end{figure}

A key property of any MCMC estimator is that statistical errors should decrease as $1/\sqrt{N}$ with the number of samples $N$. We verify this scaling at the critical temperature $\beta = \beta_c $, where sampling is challenging due to long-range correlations. Figure~\ref{fig:convergence} shows how errors in free energy, energy, and specific heat decrease with the number of MCMC sweeps. For all three quantities, the statistical error (estimated via block averaging or jackknife resampling) closely follows the expected $1/\sqrt{N}$ reference line, with scaling exponents of approximately $-0.5$. The actual errors (computed against exact values) remain at or below the statistical error estimates.

\newpage

\section{Benchmark Data Table for the $10 \times 10$ ferromagnetic Ising model}
\begin{table}[h]
\caption{Free energy per site for the $10 \times 10$ ferromagnetic Ising model with periodic boundary conditions. The critical point is $\beta_c \approx 0.44$.}
\label{tab:10x10}
\begin{ruledtabular}
\begin{tabular}{ccc}
$\beta$ & $F_{\mathrm{BP}}/N$ & $F_{\mathrm{BPLMC}}/N$ \\
\hline
0.25 & $-0.755$ & $-0.759$ \\
0.30 & $-0.784$ & $-0.793$ \\
0.34 & $-0.810$ & $-0.826$ \\
0.40 & $-0.849$ & $-0.882$ \\
0.44 & $-0.881$ & $-0.937$ \\
0.50 & $-0.933$ & $-1.033$ \\
0.59 & $-1.021$ & $-1.176$ \\
0.77 & $-1.234$ & $-1.547$ \\
1.00 & $-1.561$ & $-1.995$ \\
1.43 & $-2.276$ & $-2.862$ \\
\end{tabular}
\end{ruledtabular}
\end{table}

%======================================================================